\documentclass[conference]{IEEEtran}
\IEEEoverridecommandlockouts
\usepackage{cite}
\usepackage{amsmath,amssymb,amsfonts}
\usepackage{algorithmic}
\usepackage{graphicx}
\usepackage{textcomp}

\def\BibTeX{{\rm B\kern-.05em{\sc i\kern-.025em b}\kern-.08em
    T\kern-.1667em\lower.7ex\hbox{E}\kern-.125emX}}

\usepackage{listings}
\usepackage{xcolor}

\definecolor{codegreen}{rgb}{0,0.6,0}
\definecolor{codegray}{rgb}{0.5,0.5,0.5}
\definecolor{codepurple}{rgb}{0.58,0,0.82}
\definecolor{backcolour}{rgb}{0.95,0.95,0.92}
\lstdefinestyle{mystyle}{
  backgroundcolor=\color{backcolour},   commentstyle=\color{codegreen},
  keywordstyle=\color{magenta},
  numberstyle=\tiny\color{codegray},
  stringstyle=\color{codepurple},
  basicstyle=\ttfamily\footnotesize,
  breakatwhitespace=false,         
  breaklines=true,                 
  captionpos=b,                    
  keepspaces=true,                 
  numbers=left,                    
  numbersep=5pt,                  
  showspaces=false,                
  showstringspaces=false,
  showtabs=false,                  
  tabsize=2
}
\usepackage{fancyvrb}

\begin{document}

\title{\Huge{The Impact of DoS Attacks on Resource-constrained IoT Devices: \\A Study on the Mirai Attack}}
\author{\IEEEauthorblockN{Bhagyashri Tushir, Hetesh Sehgal, Rohan Nair, Behnam Dezfouli, Yuhong Liu }
\IEEEauthorblockA{Internet of Things Research Lab, Computer Engineering, Santa Clara University, USA\\
Emails:\{btushir, hsehgal, rnair, bdezfouli, yhliu\}@scu.edu
}
 }
 
\maketitle

\begin{abstract}
Mirai is a type of malware that creates a botnet of internet-connected devices, which can later be used to infect other devices or servers. This paper aims to analyze and explain the Mirai code and create a low-cost simulation environment to aid in the dynamic analysis of Mirai. Further, we perform controlled Denial-of-Service attacks while measuring resource consumption on resource-constrained compromised and victim Internet-of-Things (IoT) devices, such as energy consumption, CPU utilization, memory utilization, Ethernet input/output performance, and Secure Digital card usage. The experimental setup shows that when a compromised device sends a User Datagram Protocol (UDP) flood, it consumes 38.44\% more energy than its regular usage. In the case of Secure Digital usage, the victim, when flooded with Transmission Control Protocol (TCP) messages, uses 64.6\% more storage for reading and 55.45\% more for writing. The significant extra resource consumption caused by Mirai attacks on resource-constrained IoT devices can severely threaten such devices' wide adoption and raises great challenges for the security designs in the resource-constrained IoT environment.
\end{abstract}


\section{\textbf{Introduction}}
Internet of Things (IoT) is a term coined to describe any and all devices which possess the ability to communicate with each other and share information. It consists of everything from toasters, to toothbrushes, to jet engines, and operates on the concept of connectivity and communication between these devices. IoT allows for virtually endless opportunities and connections to be established and for the increased efficiency and optimization of even the simplest and most mundane tasks. For these reasons, IoT is poised to change the world. However, like anything with the Internet connectivity, IoT devices are prone to hacking, making security a paramount concern. Distributed Denial of Service (DDoS) attacks, in which multiple compromised systems are used to target a single system -- typically a server or website -- and slow or even crash it, is one of the primary attacks endangering IoT today. 

The vulnerability of IoT devices to such breaches have been demonstrated by multiple attacks recently. In September 2016, there was a DDoS attack against the KrebsOnSecurity website \cite{Ref:1}, which exceeded 600 Gigabits of traffic per second (Gbps) in volume. Following this, there was an attack on hosting company OVH \cite{Ref:2} which reached one Terabit per second (Tbps). Closely on its heels, an attack was made on Domain Name Server (DNS) provider Dyn \cite{Ref:3}, because of which websites like Twitter, Reddit, Github, and Netflix were all rendered inaccessible. These attacks were carried out by hundreds of thousands of IoT devices which were compromised by a botnet known as ‘Mirai’ \cite{Ref:11}, a type of DDoS attack in which the owner has no knowledge that his/her device has been compromised and belongs to a botnet. 

Due to the lack of security at the edge devices, the advent of IoT has created the ‘perfect storm’, a playground for malicious hackers and cyber criminals. For example, IoT edge devices are often produced for cost-efficient purposes without serious security considerations, making them very susceptible to attacks. In many cases, the owners of the IoT edge devices do not even change the default user name and password. Additionally, these edge devices do not always receive software or security updates, which can potentially result in major security breaches. However, many IoT devices have serious resource limitations, making it very challenging to directly execute standard security protocols. 

Existing studies have performed static and dynamic analyses, and defenses against Mirai in general. But most of the existing dynamic analyses are done in virtual environments, which makes it challenging to accurately measure the compromised devices' resource consumption, especially the energy consumption. To address this issue, in this work, we conduct our experiments in a controlled environment isolated from the Internet using real world devices. With a cost-efficient experimental set up, we are able to collect sufficient data on the impact of Mirai botnet on resource-constrained IoT edge devices. Specifically, Transmission Control Protocol (TCP) and User Data gram Protocol (UDP) attacks are launched from the compromised device (i.e. the device been hijacked and controlled by the attacker) towards the victim device (i.e. the device receiving malicious traffics from the compromised device). The energy consumption, processor and memory utilization, Ethernet Input Output (Ethernet IO) performance and Secure Digital card (SD card) usage on both the compromised device and the victim device are collected with and without the attacks. 

The remainder of this paper is structured as follows: Section II presents the related work and background for this work, Section III details the static analysis of Mirai attack, Section IV describes the cost-efficient experimental set up for Mirai, Section V introduces the experimental results, and Section VI finally concludes this work and proposes future directions.

\section{\textbf{Related Work and Background}}\label{ref:related_work}
As a result of the poor security, countless attacks have been made on IoT devices in the past \cite{Ref:5}, some of which have even endangered human lives. Both industry and academia are now driven to make IoT security a high priority. An analysis of IoT attack surfaces, threat models, security issues, requirements, forensics, and challenges is presented in \cite{Ref:8}. In \cite{Ref:9} the authors have analyzed the increasing threats against IoT devices and have observed that a number of attacks that used Telnet have increased since 2014. As a solution, they have used IoT honeypots -- a type of security control for capturing and analyzing malicious network traffic -- to analyze Telnet-based attacks against various IoT devices running on different CPU architectures. 

In \cite{Ref:16} the authors present an in-depth empirical security analysis of Samsung-owned smart home devices. They discovered the over-privilege problem: the analysis showed that a Smart App can have full access to the IoT devices and  battery of the app can even monitor the subscribe door lock PIN change event. In \cite{Ref:17} the authors proposed a context-based permission system, ContexIoT, for IoT platforms. ContexIoT provided contextual integrity by supporting fine-grained context identification for sensitive actions and run time prompts with rich context information that will help users to perform the access control. In \cite{Ref:18} the authors presented a large- scale analysis of firmware images. Without performing sophisticated static analysis, they discovered a total of 38 previously unknown vulnerabilities in over 693 firmware images.

Many static analysis have been conducted on Mirai \cite{Ref:10, Ref:11, Ref:12}. In \cite{Ref:10}, the authors discussed the rise of Mirai and the susceptibility of the IoT ecosystem it threatened. They also presented measurements of the botnet's evolution and DDoS activities since 2016. In \cite{Ref:13}, the authors outlined a defense strategy for Mirai attacks, but it required human involvement to reboot the device after a certain amount of time. In \cite{Ref:16} the authors explained in details IoT botnets as well as their basic modes of operation. They also discussed the major DDoS attacks including Mirai, on IoT botnets with the corresponding exploited vulnerabilities.

None of the existing work studied the dynamic analysis of Mirai in a controlled environment and the impact of Mirai on the resource consumption of the resource-constrained IoT edge devices. To complement the in-depth understanding of DDoS attacks, such as Mirai, in resource-constrained IoT environment, in this work, we study the impact of Mirai on IoT edge devices in a controlled environment with cost-efficient setup. Specifically, the memory usage, processor utilization, energy consumption, SD card usage, and Ethernet IO performance on both the compromised and victim devices are collected and analyzed. 

\section{\textbf{Static Analysis of Mirai}}
This section will briefly discuss the design of Mirai, which aims to create a botnet of infected devices to execute DDoS attacks. In particular, Mirai targets a specific type of vulnerable IoT and embedded devices, such as routers, IP cameras, Digital Video Recorder (DVRs), and printers \cite{Ref:10}, because these devices often have very few build-in security protections and sometimes do not even require a user name and password. 

Mirai source code was leaked, making its analysis feasible \cite{Ref:15}. Generally speaking, Mirai works by first carrying out the scanning phase, in which it pseudo-randomly generates IPv4 addresses and sends TCP SYN requests on Telnet port 23 and 2323. If there is a response from a device, the botnet goes into a brute force login phase in which it tries 62 different pre configured credentials. After a successful login, Mirai stores the compromised device's IP address in a dedicated report server. The malware then determines the underlying environment of the compromised device, downloads and executes an architecture-specific malware. At this point, the attacker can send attack commands to the botnet while the bots are scanning for other vulnerable devices. 

In particular, the source code consists of three component: bot, CNC server, and loader. 

\textbf{Bot}: The Bot is written in C and runs on the compromised IoT devices. Once executed on the RAM of the compromised device, this branch of Mirai deletes its own file to avoid detection, and disables the watchdog timer which prevents the compromised device from rebooting. There are three modules running besides the main process: attack, killer and scanner. 

\begin{itemize}
\item Attack: This module commences with the DDoS attack when it receives the command from the attacker and terminates once its duration time expires. Ten different DoS attacks can be launched by Mirai. 

\item Killer: This module kills processes using ports 22, 23 and 80, and then reserves these ports. It continuously scans memory to find and kill any malware deployed by other attacks.
 
\item  Scanner: This module uses telnet protocol to generate random IPv4 addresses. If a device is compromised, its IP address will be sent to the reporting server with its default user name and password. 
\end{itemize}

\textbf{CNC server}: The CNC server is written in GO and runs remotely (e.g. in a cloud). There are two sockets in this component: one taking port 23 and the other taking port 101. Port 23 is used by a new bot to make connection with CNC and port 101 is used by registered bots. By using predefined credentials, the bots can connect to a MySQL database. Upon connection, the CNC server decides if the connection request is from CNC registered bots or from a new bot, in which case it is added to the list.

\textbf{Loader}: Loader is written in C. To pre-compile payload for different architectures, this component first creates a server and then listens to vulnerable IoT devices. After receiving the information about the compromised device, loader connects to it via the Telnet protocol and runs the payloads based on specific architectures.

\section{\textbf{System Design: Mirai Experiment Setup}}

This section describes a cost-efficient lab set up to simulate TCP and UDP attacks from a Mirai compromised device. This experiment focuses on collecting incoming malicious network traffic and the usage statistics of both a compromised device and a victim device. A compromised device refers to the device that is hijacked by the Mirai virus and used to launch a botnet attack. On the other hand, a victim device refers to the device that is attacked by the compromised device. Our test bed is cost-efficient, as it includes only two raspberry Pi's, and one Ethernet cable. Another Raspberry Pi is optional, in case one also wants to measure the energy consumption of devices under test. The prices of the devices used for the experiment are given in Table \ref{Table.1}. 
\begin{figure}[tb]
\centering
\includegraphics[width=3in]{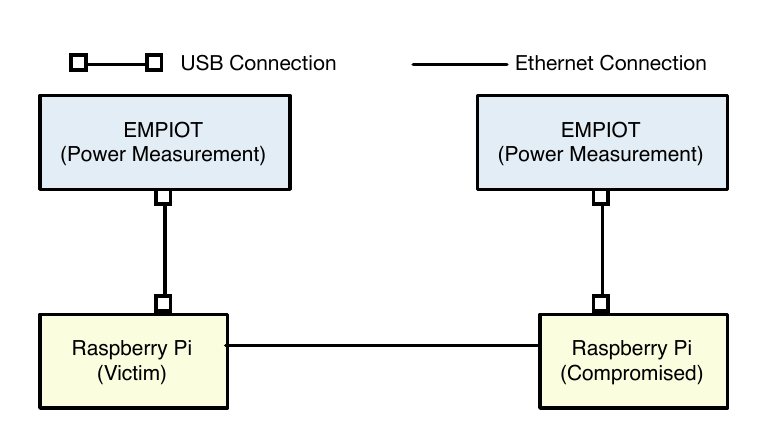}
\caption{Components of testbed and interconnection.} \label{fig:exp_setup}
\setlength{\belowcaptionskip}{-1000pt}
\end{figure}
 \vspace{-0.10 cm}
\begin{table}[!tb] 
 \centering
\def\arraystretch{1.2}
\caption{Devices Used for Experimentation Set Up}
\begin{tabular} { |p{2.5cm}|p{1.5cm}|p{1.5cm}| }
 \hline
 Device             & Price (USD)  & CPU/Mem \\   \hline \hline

 Raspberry Pi 3 (two)  & 159.98    &   1.2GHZ (1GB)    \\  \hline

 Empiot                 &  79.99    &    1.2GHZ (1GB) \\  \hline

 Ethernet Cable     &   6.9         &         -                \\ \hline
  
 \end{tabular}
\label{Table.1}
\end{table}

Figure \ref{fig:exp_setup} depicts the topology used in our experiment. One Raspberry Pi will act as the compromised device, while the other will act as the victim. The compromised Raspberry Pi will send DOS attacks to the victim Raspberry Pi, via an Ethernet connection. The first objective is to capture the incoming packets from the compromised device. {\fontfamily{pcr}\selectfont Iperf3}, a free binary available online, is used to simulate a flood of incoming TCP and UDP DoS packets. The binary is setup on both the victim and compromised device. A server to receive the packets is set up on the victim device through the following command:
 \begin{center}
    {\fontfamily{pcr}\selectfont iperf3 -s}
\end{center}
The compromised device then prepares to send a flood of UDP packets using the command:
\vspace{1cm}
\begin{lstlisting}[caption={Bash Script to measure Mirai statistics},label={lst:label}]
!/bin/bash
# Continue to measure usage statistics until force quit
while true; do
    # Save current time to output file
    date +%r >> output.txt

    # Save idle CPU usage using mpstat command
    mpstat | grep -Po 'all.* \K[^ ]+$' >> output.txt
    
    #Print memory usage using free mem command 
    free | grep Mem | awk '{print $3/$2 * 100.0}' >> output.txt 

    #Save SD read/write statistics using iostat command 
    iostat -dx mmcblk0 | grep mmcblk0 | awk '{ print $6"/"$7; }' >> output.txt 
    
    #Save Ethernet read/write statistics using ifstat command (Kbps) 
    ifstat -i eth0 0.1 1 | awk 'FNR == 3 {print $1"/"$2;}'>> output.txt 
    
    #Sleep for 100 milliseconds 
    printf "\n" >> output.txt sleep 0.1 
done
\end{lstlisting}

\begin{center}
{\fontfamily{pcr}\selectfont./iperf3 -u -c $<$ IP of Victim $>$ -b 30M -l 1200 -P 6 -t 60}
\end{center}

This command sends packets with a rate of 30 Mega bits per second (Mbps), a buffer length of 1200 bits for a duration of 60 seconds. 
Using a package called {\fontfamily{pcr}\selectfont t-shark}, packet capturing is prepared on the victim device with the following command:
\begin{center}
{\fontfamily{pcr}\selectfont tshark -i eth0  host $<$IP of ethernet cable$>$ $>>$ udp.txt.}
\end{center}

This ensures that packets coming over the Ethernet cable with the specified IP are captured and stored in \texttt{udp.txt}. The command to send UDP packets is then executed on the compromised device, alongside the command to capture the packets on the victim device. Same steps are then repeated to capture incoming TCP packets, which are sent using the following command:
\begin{center}
{\fontfamily{pcr}\selectfont./iperf3 -c $<$IP of Victim$>$ -b 30M -l 1200 -P 6 -t 60.}
\end{center}

After completing the packet collection, we next capture the usage statistics from both the compromised device and the victim device. To measure the Raspberry Pis' energy consumption, a separate device named the Empiot \cite {dezfouli2018empiot} is used. The Empiot is a low-cost, easy-to-build, and flexible energy measurement platform. It takes 1000 samples per second and has an accuracy of 100 mA. The script to measure the CPU usage, memory usage, SD usage, and Ethernet I/O runs on both the victim and compromised devices. The listed usage statistics are measured using the bash script shown in Listing \ref{lst:label}.

The attacks are sent using the exact same TCP and UDP flood commands during packet collection. Using the topology depicted by Figure \ref{fig:exp_setup}, the malicious TCP and UDP traffic are separately sent to the victim device, while all usage statistics are being recorded on the compromised and victim device. Each attack is simulated for a duration of 60 seconds, and all usage statistics are also recorded for the same duration. 

In order to account for the added overhead by measurement code, additional recordings for the usage are made with just the bash script running on both the victim and compromised Pi's. During this period, no attacks are sent. These results are then subtracted from the results of our experiment to account for the extra overhead caused by the attack. We observe that normalized results are only 1\% lower, which indicates that our bash scripts are also very efficient. Further, even with a low cost setup, we are able to demonstrate that the system can be overwhelmed by a Mirai DoS attack.
\begin{figure}[tb]
\centering
\includegraphics[width=3.5in]{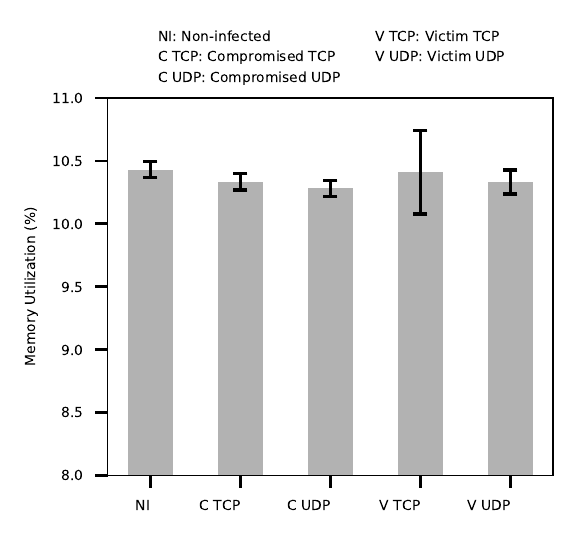}
\caption{Memory utilization of various devices under DoS attack.}\label{fig:mem}
\vspace{-0.10 cm}
\setlength{\belowcaptionskip}{-60pt}
\end{figure}

\begin{figure}[tb]
\centering
\includegraphics[width=3.5in]{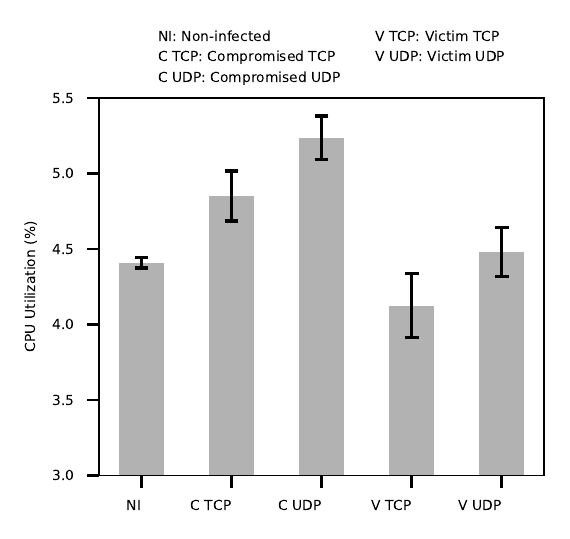}
\caption{CPU utilization of various devices under DoS attack.}\label{fig:cpu}
\vspace{-0.10 cm}
\setlength{\belowcaptionskip}{-60pt}
\end{figure}

\section{\textbf{Performance Evaluation}}
\begin{figure}[tb]
\centering
\includegraphics[width=3.5in]{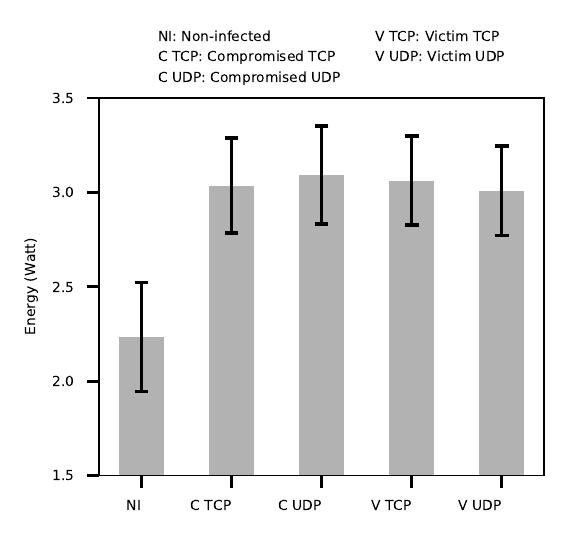}
\caption{Energy consumption of various devices under DoS attack.}\label{fig:energy}
\vspace{-0.10 cm}
\setlength{\belowcaptionskip}{-60pt}
\end{figure}

\begin{figure}[tb]
\centering
\includegraphics[width=3.5in]{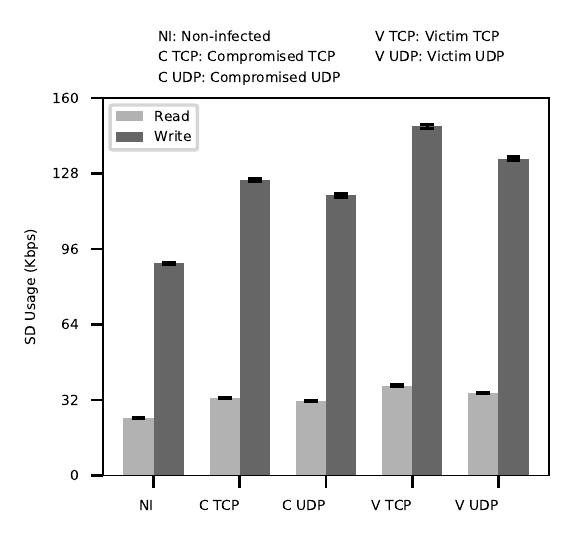}
\caption{SD card utilization of various devices under DoS attack.}\label{fig:sd}
\vspace{-0.10 cm}
\setlength{\belowcaptionskip}{-60pt}
\end{figure}

\begin{figure}[tb]
\centering
\includegraphics[ width=3.5in]{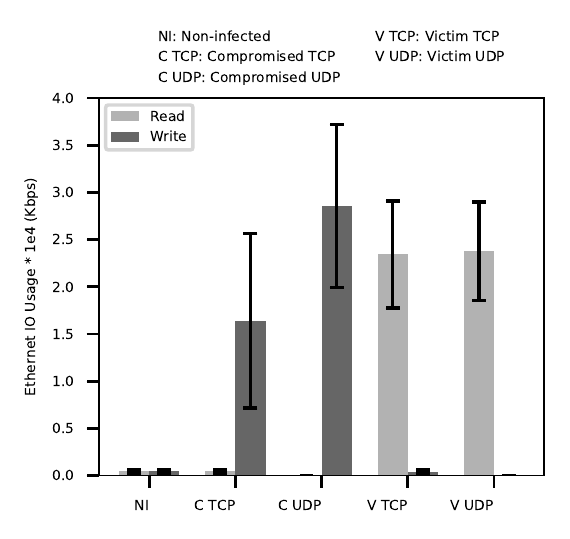}
\caption{Ethernet IO utilization of various devices under DoS attack.}\label{fig:ethio}
\vspace{-0.10 cm}
\setlength{\belowcaptionskip}{-60pt}
\end{figure}

In this section, we compare the resource consumption of a compromised and victim device to that of a normal device. In particular, memory utilization, CPU utilization, energy consumption, SD card usage and Ethernet IO usage are illustrated in Figures \ref{fig:mem}, \ref{fig:cpu}, \ref{fig:energy}, \ref{fig:sd}, and \ref{fig:ethio}. In each Figure, the x-axis represents different devices, namely non-infected, compromised TCP, compromised UDP, victim TCP, and victim UDP. 
\begin{itemize}
    \item \textbf{Non-infected}: In this scenario, no attack is launched on the device.
    \item \textbf{Compromised TCP}: The device is compromised and launching a TCP flood attack against another device.
    \item \textbf{Compromised UDP}: The device is compromised and launching a UDP flood attack against another device.
    \item \textbf{Victim TCP}: The device is a victim device and flooded by malicious TCP packets.
    \item \textbf{Victim UDP}: The device is a victim device and flooded by malicious UDP packets.
\end{itemize}

Figure \ref{fig:mem} compares the memory utilization measured in percentages for different devices. We can observe compared to a non-infected device, the memory utilization decreases by 0.93\% when a compromised device conducts a TCP flood, 0.2\% when a victim device receives a TCP flood, 1.44\% when a compromised device launches a UDP flood, and 0.96\% when a victim device receives a UDP flood. The decrease in memory utilization is because of the Raspberry Pi’s computing constraints. Given that the Raspberry Pi is flooded by the memory read/write tasks, it does not have enough space to handle simultaneously multitasks -- processing incoming/outgoing packets and writing/reading them to/from memory, leading to lower memory utilization. 

Next, comparing compromised TCP to compromised UDP and victim TCP  to victim UDP, the memory utilization is more for the victim and compromised TCP devices. This is because TCP needs more memory to keep track of per-connection states, for example, sequence numbers, buffer for out-of-order data, and acknowledgments. Further, we can observe that compromised TCP and compromised UDP devices consume less memory than victim TCP and victim UDP devices, respectively. This validates victim devices spend more memory in processing the packets rather than receiving them. The reason is that the victim device does not receive all the packets as confirmed by the tshark file. 

Figure \ref{fig:cpu} compares CPU utilization measured by percentages for different devices. As anticipated, compared to a non-infected device, the compromised devices launching the TCP and UDP floods utilize 10.05\% and 18.77\% more CPU, respectively. However, as opposed to the non-infected device, the victim devices under a TCP and UDP flood use less CPU by 6.42\% and 1.6\%, respectively. This is because packets flood the victim's CPU while the victim has to read and process the flood packets simultaneously. However, given that Raspberry Pi has limited resources, it cannot do both. Next, comparing TCP compromised to UDP compromised and TCP victim to UDP victim, we can see UDP victim and compromised utilizes more CPU. From the tshark files, we observe the number of TCP packets received by the victim device is less than the number of UDP packets. This is because TCP applies congestion control. Finally, compromised TCP and compromised UDP devices utilize more CPU than victim TCP and victim UDP devices, respectively. This is because the victim device does not receive all the packets sent by the compromised device. Thus, less CPU utilization. 

Figure \ref{fig:energy} compares the energy usage measured in watt for different devices. As expected, compared to a non-infected device, the energy consumption is increased by 35.87\% when a device is compromised and performs a TCP flood, 37.04\% when a victim receives a TCP flood, 38.44\% when a device is compromised and launches a UDP flood, and 34.61\% when a victim receives a UDP flood. The increase in energy consumption is due to the extra packets to be handled by the compromised device or the victim device. Next, comparing compromised TCP to compromised UDP, the latter consumes more energy. This is because compromised TCP sends fewer packets. However, comparing victim TCP to victim UDP, victim TCP consumes more energy. Even though the victim TCP receives fewer packets than the victim UDP, it consumes more memory. Thus, higher energy consumption. Lastly, compromised TCP consumes less energy than victim TCP. This is because victim TCP continuously processes the packets, but compromised TCP does not send packets always since it applies congestion control. In contrast, compromised UDP consumes more energy than victim UDP since the number of packets received is less than sent.

Figure \ref{fig:sd} compares the read/write cycles of the SD card measured in Kilobits per second (Kbps). For both the compromised device and the victim device, it is observed that the number of writes is less than the number of reads. This makes sense as after reading the incoming TCP or UDP packets from a flood, the victim device discards them, instead of storing them in memory. In particular, when a compromised device is bombarded with a TCP flood, 39.22\% more reads and 34.68\% more writes are performed compared to the normal device. A victim device, under the TCP flood, performs 64.6\% more reads and 55.45\% more writes. In the case of the UDP flood, a compromised device performs 32.03\% more reads and 29.22\% more writes and a victim device performs 49.38\% more reads and 42.98\% more writes.

Figure \ref{fig:ethio} compares the Ethernet IO measured in Kilobits per second (Kbps) for different devices. As minimum information is being sent or received via Ethernet by the non-infected device, it yields zero reads and writes. Subsequently, comparing compromised TCP to compromised UDP, the Ethernet writes are more for compromised UDP because the compromised UDP device sends more packets are than compromised TCP. On a similar note, victim UDP devices read more packets than victim TCP packets. Further, we can observe that compromised TCP and compromised UDP has more writes than victim TCP and victim UDP devices, respectively. This is because compromised devices send (i.e., writes) packets, and the victim device receives (i.e., reads) the packets. 

\section{ Conclusion and Future work}
Mirai is a malware that turns a compromised device into a bot. This bot becomes part of a botnet, which later can be used to execute DDoS attacks. Based on its history, Mirai is among one of the highest profile attacks, making its analysis significant. In this work, we conduct both static and dynamic analysis of Mirai. In particular, we have conducted all the experiments in an isolated and cost-efficient experimental setup. It is observed that Mirai has a significant impact on the resource consumption of IoT edge devices. When an IoT edge device is flooded with TCP and UDP packets, there are significant increases in the usage of the processor, Ethernet IO, SD usage and decreases in memory utilization. The analysis of such resource consumption will benefit the deep understanding of DDoS attacks' impact on resource-constrained IoT environments and facilitate future researches on light-weight defense mechanisms against such attacks. 
\bibliography{references.bib}

\begin{thebibliography}{10}
\providecommand{\url}[1]{#1}
\csname url@samestyle\endcsname
\providecommand{\newblock}{\relax}
\providecommand{\bibinfo}[2]{#2}
\providecommand{\BIBentrySTDinterwordspacing}{\spaceskip=0pt\relax}
\providecommand{\BIBentryALTinterwordstretchfactor}{4}
\providecommand{\BIBentryALTinterwordspacing}{\spaceskip=\fontdimen2\font plus
\BIBentryALTinterwordstretchfactor\fontdimen3\font minus
  \fontdimen4\font\relax}
\providecommand{\BIBforeignlanguage}[2]{{%
\expandafter\ifx\csname l@#1\endcsname\relax
\typeout{** WARNING: IEEEtran.bst: No hyphenation pattern has been}%
\typeout{** loaded for the language `#1'. Using the pattern for}%
\typeout{** the default language instead.}%
\else
\language=\csname l@#1\endcsname
\fi
#2}}
\providecommand{\BIBdecl}{\relax}
\BIBdecl

\bibitem{Ref:1}
B.~Krebs. Krebsonsecurity hit with record ddos. [online]. Available:
  https://krebsonsecurity.com/2016/09/krebsonsecurity-hit-with-record-ddos/.

\bibitem{Ref:2}
O.~Klaba. (2016) Octave klaba [online]. Available:
  https://twitter.com/olesovhcom/status/778830571677978624.

\bibitem{Ref:3}
S.~H. (2016). Dyn analysis summary of friday october 21 attack. [online].
  Available:
  https://dyn.com/blog/dyn-analysis-summary-of-friday-october-21-attack/.

\bibitem{Ref:11}
C.~Kolias, G.~Kambourakis, A.~Stavrou, and J.~Voas, ``Ddos in the iot: Mirai
  and other botnets,'' \emph{Computer}, vol.~50, no.~7, pp. 80--84, 2017.

\bibitem{Ref:5}
T.~G. (2016). Team of hackers take remote control of tesla model from 12 miles
  away. [online]. Available:
  https://dyn.com/blog/dyn-analysis-summary-of-friday-october-21-attack/.

\bibitem{Ref:8}
M.~M. Hossain, M.~Fotouhi, and R.~Hasan, ``Towards an analysis of security
  issues, challenges, and open problems in the internet of things,'' in
  \emph{IEEE World Congress on Services (SERVICES)}, 2015, pp. 21--28.

\bibitem{Ref:9}
\BIBentryALTinterwordspacing
Y.~M.~P. Pa, S.~Suzuki, K.~Yoshioka, T.~Matsumoto, T.~Kasama, and C.~Rossow,
  ``Iotpot: Analysing the rise of iot compromises,'' in \emph{9th {USENIX}
  Workshop on Offensive Technologies ({WOOT} 15)}.\hskip 1em plus 0.5em minus
  0.4em\relax Washington, D.C.: {USENIX} Association, 2015. [Online].
  Available:
  \url{https://www.usenix.org/conference/woot15/workshop-program/presentation/pa}
\BIBentrySTDinterwordspacing

\bibitem{Ref:16}
E.~Fernandes, J.~Jung, and A.~Prakash, ``Security analysis of emerging smart
  home applications,'' in \emph{Security and Privacy (SP), IEEE Symposium
  on}.\hskip 1em plus 0.5em minus 0.4em\relax IEEE, 2016, pp. 636--654.

\bibitem{Ref:17}
Y.~J. Jia, Q.~A. Chen, S.~Wang, A.~Rahmati, E.~Fernandes, Z.~M. Mao,
  A.~Prakash, and S.~J. Unviersity, ``Contexiot: Towards providing contextual
  integrity to appified iot platforms,'' in \emph{Proceedings of The Network
  and Distributed System Security Symposium}, 2017.

\bibitem{Ref:18}
A.~Costin, J.~Zaddach, A.~Francillon, D.~Balzarotti, and S.~Antipolis, ``A
  large-scale analysis of the security of embedded firmwares.'' in \emph{USENIX
  Security Symposium}, 2014, pp. 95--110.

\bibitem{Ref:10}
M.~Antonakakis, T.~April, M.~Bailey, M.~Bernhard, E.~Bursztein, J.~Cochran,
  Z.~Durumeric, J.~A. Halderman, L.~Invernizzi, M.~Kallitsis \emph{et~al.},
  ``Understanding the mirai botnet,'' in \emph{USENIX Security Symposium},
  2017.

\bibitem{Ref:12}
D.~Web, ``Investigation of linux mirai trojan family. [online],'' available:
  https://www.yumpu.com/en/document/view/55993587/investigation-of-linuxmirai-trojan-family.

\bibitem{Ref:13}
Cao, ``Hey, you, keep away from my device: remotely implanting a virus expeller
  to defeat mirai on iot devices,'' \emph{arXiv preprint arXiv:1706.05779},
  2017.

\bibitem{Ref:15}
Anna-senpa. (2016) Mirai-source-code. [online]. Available:
  https://github.com/jgamblin/Mirai-Source-Code.

\bibitem{dezfouli2018empiot}
B.~Dezfouli, I.~Amirtharaj, and C.-C. Li, ``Empiot: An energy measurement
  platform for wireless iot devices,'' \emph{arXiv preprint arXiv:1804.04794},
  2018.

\end{thebibliography}
\bibliographystyle{IEEEtran}
\end{document}